\documentclass[12pt]{iopart}
\input{psfig.sty}

\newcommand*{\rf}[1]{Fig. \ref{fig:#1}}
\newcommand*{\re}[1]{Eq. \ref{eq:#1}}

\def\cmm2{{\,\rm cm^{-2}}}
\def\cm2{{\,{\rm cm}^2}}
\def\cmm3{{\,{\rm cm}^{-3}}}
\def\gcmm3{{\,{\rm g\,cm^{-3}}}}

\newcommand{\bmn}{\mathbf{n}}
\newcommand{\bmk}{\mathbf{k}}
\newcommand{\bmx}{\mathbf{x}}
\newcommand{\bml}{\mathbf{l}}
\newcommand{\bmo}{\mathbf{0}}

\def\VEV#1{\left\langle #1\right\rangle}
\def\fun#1#2{\lower3.6pt\vbox{\baselineskip0pt\lineskip.9pt
  \ialign{$\mathsurround=0pt#1\hfil##\hfil$\crcr#2\crcr\sim\crcr}}}

\def\be{\begin{equation}}
\def\ee{\end{equation}}
\def\bea{\begin{eqnarray}}
\def\eea{\end{eqnarray}}

\makeatletter
\newcommand{\contraction}[5][1ex]{%
  \mathchoice
    {\contraction@\displaystyle{#2}{#3}{#4}{#5}{#1}}%
    {\contraction@\textstyle{#2}{#3}{#4}{#5}{#1}}%
    {\contraction@\scriptstyle{#2}{#3}{#4}{#5}{#1}}%
    {\contraction@\scriptscriptstyle{#2}{#3}{#4}{#5}{#1}}}%
\newcommand{\contraction@}[6]{%
  \setbox0=\hbox{$#1#2$}%
  \setbox2=\hbox{$#1#3$}%
  \setbox4=\hbox{$#1#4$}%
  \setbox6=\hbox{$#1#5$}%
  \dimen0=\wd2%
  \advance\dimen0 by \wd6%
  \divide\dimen0 by 2%
  \advance\dimen0 by \wd4%
  \vbox{%
    \hbox to 0pt{%
      \kern \wd0%
      \kern 0.5\wd2%
      \contraction@@{\dimen0}{#6}%
      \hss}%
    \vskip 0.5ex
    \vskip\ht2}}

\newcommand{\contraction@@}[3][0.05em]{%
  \hbox{%
    \vrule width #1 height 0pt depth #3%
    \vrule width #2 height 0pt depth #1%
    \vrule width #1 height 0pt depth #3%
    \relax}}
\makeatother

\begin{document}
\title[Second-order Weak Lensing Corrections]{The Born and Lens-Lens Corrections to Weak Gravitational Lensing Angular Power Spectra}

\author{Charles Shapiro}
\address{Department of Physics, The
University of Chicago, Chicago, IL~~60637}
\address{Kavli Institute for Cosmological Physics, The
University of Chicago, Chicago, IL~~60637}

\author{Asantha Cooray}
\address{Center for Cosmology, Department of Physics and Astronomy, University of California, Irvine, CA~~92617}

\date{\today}

\begin{abstract}
We revisit the estimation of higher order corrections to the angular power spectra of weak gravitational lensing. Extending a previous calculation of Cooray and Hu, we find two additional terms to the fourth order in potential perturbations of large-scale structure corresponding to corrections associated with the Born approximation and the neglect of line-of-sight coupling of two foreground lenses in the standard first order result. These terms alter the convergence ($\kappa\kappa$), the lensing shear E-mode ($\epsilon\epsilon$), and their cross-correlation ($\kappa\epsilon$) power spectra on large angular scales, but leave the power spectra of the lensing shear B-mode ($\beta\beta$) and rotational ($\omega\omega$) component unchanged as compared to previous estimates. The new terms complete the calculation of corrections to weak lensing angular power spectra associated with both the Born approximation and the lens-lens coupling to an order in which the contributions are most significant. Taking these features together, we find that these corrections are unimportant for any weak lensing survey, including for a full sky survey limited by cosmic variance.
\end{abstract}

\maketitle

\section{Introduction}

Weak gravitational lensing provides a powerful way to probe the matter distribution of the large-scale structure of the Universe and 
to  measure cosmological parameters both through growth of structures and geometrical distance projections (see, \cite{Bar01,Ref,Sch05} for recent reviews).
As weak lensing experiments improve in precision, it is necessary to improve the accuracy to which lensing statistics are
computed.  Included in the sources of theoretical uncertainties are various approximations used in the canonical calculation of the weak lensing power spectra.  
Two such approximations, the Born approximation and the exclusion of lens-lens coupling involving two foreground lenses
along the line-of-sight, were investigated by Cooray and Hu \cite{CH} using analytical methods, while
numerical simulations have also been used to explore these approximations \cite{Jain99,ValeWhi03}.
Relaxing these approximations introduce sub-percent-level corrections to the power spectra of the convergence 
($\kappa$) and the shear E-mode ($\epsilon$).  
Miniscule contributions to the angular power spectra of shear B-mode ($\beta$) and image rotational ($\omega$) components are also generated, though
in the standard calculation, these two components are exactly zero. This has led to the use of the shear B-mode component 
monitoring systematics in lensing studies.

In Cooray and Hu \cite{CH}, the estimation of lensing corrections was done perturbatively; they computed corrections to the power spectra which were fourth order in the gravitational potential of large-scale structure.  
In this paper, we briefly review these corrections and show that there exist two additional terms of the same order.  Mainly, these terms decrease the power of the convergence and shear $\epsilon$-mode on large angular scales.  
Neither the shear $\beta$-mode nor the rotational component are changed relative to the previous estimate in Ref.~\cite{CH}.   Together, the calculation is complete
and we can make a strong conclusion that the
corrections due to the Born approximation and lens-lens coupling are negligible for both current and upcoming lensing experiments. Unless
one begins to probe multipole moments greater than $10^4$, it is likely that these corrections can be ignored for any experiment, including an ultimate
all-sky experiment limited by cosmic variance alone. While the calculation related to higher order corrections is complete, there may be other
sources of error in the theoretical estimates of lensing statistics \cite{Ref,Jain05}. 
For example, theoretical estimates of the underlying non-linear dark matter clustering is
not known to better than 10\% at arcminute scales \cite{Smith03} and is likely to be further complicated by baryon physics \cite{White04,Zhan04}.
Another source of important error is that lensing statistics are based on reduced shear rather than shear directly, which results in
perturbative corrections from convergence to shear \cite{White05}. 
Clustering of background sources also contaminates weak lensing
 statistics \cite{cluster} and may be an important source of uncertainty for tomographic lensing measurements \cite{tomog}
 in which the background source redshift distribution is binned to obtain lensing
 measurements as a function of redshift.  Based on our calculation, we are now confident that
that these remaining issues dominate the theoretical calculation and interpretation of the weak lensing angular power spectra.

This brief paper is organized as following: In the next Section, we outline the calculation related to higher order corrections to weak lensing angular power spectrum.
We identify two new terms by extending the calculation of Ref.~\cite{CH}, and make numerical estimates of the correction. We find that the overall correction is
negligible even for an all-sky lensing experiment limited by  cosmic variance and probing fluctuations out to a multipole of $10^4$ in the angular power
spectrum. We conclude with a brief summary of our results in Section~\ref{Sec:Results}. For illustrative purposes, we assume a 
flat $\Lambda$CDM cosmological model with parameters $\Omega_c=0.3$, $\Omega_b=0.05$, $\Omega_\Lambda=0.65$
for CDM, baryon, and cosmological constant densities relative to the closure density,
a scaled Hubble constant of $h=0.7$, and a primordial power spectrum of fluctuations normalized to $\sigma_8=0.85$ with a
tilt of $n=1$.

\section{Calculational Method}

Here we provide a brief summary of the analytical calculation. For further details, we refer the reader to Cooray \& Hu \cite{CH}.
For simplicity, we also use their notation throughout.

The distortion tensor for a weakly lensed source is given by an integral over a photon's path, $\bmx$, through the large-scale gravitational potential:
\be
\psi_{ab}(\bmn,\chi_s) = 2\int_0^{\chi_S} d\chi\; g(\chi,\chi_s) \Phi_{,ac}(\bmx;\chi)[\delta_{cb}+\psi_{cb}(\bmn,\chi)] , \label{eq:distort}
\ee
where $\bmn$ is the source's sky position, $\chi_s$ is its comoving distance from us, $\Phi$ is the potential and $\delta$ is the Kronecker delta.  Commas denote spatial derivatives in the transverse directions, and there is an implicit sum over repeated indices.  The weighting function $g$ assumes that the source is at a 
fixed distance away; it is given by
\be
g(\chi',\chi) \equiv
\left\{
\begin{array}{ll}
d_A(\chi-\chi') d_A(\chi')/d_A(\chi)& \mbox{ for $\chi'<\chi$} \\
0 & \mbox{ for $\chi'\ge \chi$}
\end{array}
\right.
\ee
with $d_A$ as the angular diameter distance.  The power spectra of weak lensing observables are found by combining components of $C_{abcd}$, defined as
\be
\VEV{\psi^*_{ab}(\bml)\psi_{cd}(\bml')}\equiv(2\pi)^2\delta(\bml-\bml')C_{abcd}(\bml),
\ee 
with $\psi(\bml)$ the 2D Fourier transform of $\psi(\bmn)$.

The canonical calculation of the power spectra takes advantage of the fact that the potential -- and therefore the deflection of light -- is small.  Lens-lens coupling, i.e. the appearance of $\psi$ on the right side of equation~\re{distort}, 
can clearly be ignored to first order in $\Phi$.  Furthermore, since the photon path is nearly a straight line, we may write
\be
\bmx(\bmn,\chi)=\bmn d_A(\chi) + \delta\bmx(\bmn,\chi), \label{eq:x}
\ee
and in the Born approximation set the transverse deflection $\delta\bmx$ to zero.  The result is a simplified integral over the undeflected path,
\be
\psi_{ab}(\bmn,\chi_s) = 2\int_0^{\chi_S} d\chi\; g(\chi,\chi_s) \Phi_{,ab}(\bmn d_A(\chi);\chi) + O(\Phi^2),
\ee
which, in the Limber approximation \cite{Lim54}, leads to the standard expression for the two point functions of $\psi$ in terms of the power spectrum of $\Phi$ \cite{Kai90}
\be
C_{abcd}(\bml)=4l_al_bl_cl_d\int d\chi \frac{g(\chi,\chi_S)^2}{d^6_A(\chi)}P(l/d_A;\chi) \, .
\ee
The first order spectra of the lensing observables ($\kappa, \epsilon, \beta, \omega$) are
\be
C^{\kappa\kappa}_l=C^{\kappa\epsilon}_l=C^{\epsilon\epsilon}_l=l^4\int d\chi \frac{g(\chi,\chi_S)^2}{d^6_A(\chi)}P(l/d_A;\chi)
\ee
with all other combinations equal to zero. Note that throughout this paper, we will be employing Limber approximation to
calculate angular power spectra.  This assumption makes use of the flat-sky approximation and could potentially affect the
large angular scale correlations. However, as a test case, we considered the calculation exactly by integrating over
spherical Bessel functions that one encounters with the all-sky projections and found our results to be
accurate at the percent level and below at $\ell > 10$. Just as the Limber approximation is adequate to
calculate convergence power spectrum at the first order (e.g., Ref. \cite{WhiHu00} based on
simulations), we believe it is adequate for the calculations at the higher order.

\subsection{Perturbative Corrections}

To relax the Born approximation, we simply Taylor expand the potential in equation~\re{distort} about the undeflected photon path.  With the transverse deflection given by
\be
\delta x_a(\bmn,\chi)=-2\int d\chi' g(\chi',\chi)\frac{d_A(\chi)}{d_A(\chi')}\Phi_{,a}(\bmx;\chi')\, , \label{eq:dx}
\ee
the potential is
\bea
\Phi(\bmx;\chi)&=&\Phi(\bmn d_A(\chi)+\delta\bmx;\chi) = \Phi(\bmn d_A;\chi)+\delta x_a\Phi_{,a}(\bmn d_A;\chi) \nonumber \\
&& + \frac{1}{2}\delta x_a\delta x_b\Phi_{,ab}(\bmn d_A;\chi)+O(\Phi^4)\, . \label{eq:expand}
\eea
Note that $\delta\bmx$ depends on the deflected path, and must also be expanded perturbatively.  Substituting equation~\re{expand} into equation~\re{dx}, 
we find the deflection to second order in potential fluctuations to be:
\bea
\delta x_a(\bmn,\chi)&=&\delta x_a^{(1)}(\bmn,\chi)+\delta x_a^{(2)}(\bmn,\chi)+O(\Phi^3) \\
\delta x_a^{(1)}(\bmn,\chi)&\equiv&-2\int d\chi'g(\chi',\chi)\frac{d_A(\chi)}{d_A(\chi')}\Phi_{,a}(\bmn d_A(\chi');\chi') \\
\delta x_a^{(2)}(\bmn,\chi)&\equiv&-2\int d\chi'g(\chi',\chi)\frac{d_A(\chi)}{d_A(\chi')}\Phi_{,ab}(\bmn d_A(\chi');\chi')\delta x_b^{(1)}(\bmn,\chi') \, .
\eea
Now we may express $\Phi(\bmx)$, the potential at the deflected position, using integrals over the undeflected path:
\bea
\Phi(\bmx;\chi)&=&\Phi(\bmn d_A;\chi)+(\delta x_a^{(1)}+\delta x_a^{(2)})\Phi_{,a}(\bmn d_A;\chi)\nonumber \\
&& +\frac{1}{2}\delta x_a^{(1)}\delta x_b^{(1)}\Phi_{,ab}(\bmn d_A;\chi)+O(\Phi^4)\, . \label{eq:expand3}
\eea
Inserting equation~\re{expand3} into equation~\re{distort} leads to $\psi$ as a line-of-sight projection of a source field,
\be
\psi_{ab}(\bmn,\chi_S)=2\int d\chi \, g(\chi,\chi_S)S_{ab}(\bmn d_A;\chi),
\ee
where $S_{ab}(\bmn d_A;\chi)$ includes the following terms (all depend on $\chi$):
\bea
S^{(1 )}_{ab}&=&\Phi_{,ab}(\bmn d_A) \\
S^{(2B)}_{ab}&=&\delta x_c^{(1)}(\bmn)\Phi_{,abc}(\bmn d_A) \\
S^{(3X)}_{ab}&=&\delta x_c^{(2)}(\bmn)\Phi_{,abc}(\bmn d_A) \\
S^{(3B)}_{ab}&=&(1/2)\delta x_c^{(1)}(\bmn)\delta x_d^{(1)}(\bmn)\Phi_{,abcd}(\bmn d_A) \, .
\eea
These terms yield the standard second order expression for $C_{abcd}$ plus higher order corrections.  Third order corrections vanish in the Limber approximation.  Cooray and Hu \cite{CH} computed all fourth order corrections except for a term that couples $S^{(3X)}$ to $S^{(1)}$.  We find that it is given by
\bea
C^X_{abcd}(\bml)&=&4\int d\chi\frac{g(\chi,\chi_S)^2}{d_A(\chi)^6}P^X_{abcd}(\bml;\chi) \\
P^X_{abcd}(\bml;\chi)&=&-4\int\frac{d^2l'}{(2\pi)^2}(l_al_bl'_cl'_d+l'_al'_bl_cl_d)(\bml\cdot\bml')^2\int d\chi' \frac{g(\chi',\chi)^2d_A(\chi)^2}{d_A(\chi')^8}\nonumber \\
\quad &\times& P\left({l\over d_A(\chi)};\chi\right)P\left({l'\over d_A(\chi')};\chi'\right) \, , \label{eq:PX}
\eea
where $P^X_{abcd}$ is the correction to the power spectrum of the source field.  The corrections to the power and cross spectra of the lensing observables ($\kappa,\epsilon,\beta,\omega$) are
\bea
C^{X\alpha\beta}_l&=&-{\eta\over\pi}\,l^4\int d\chi{g(\chi,\chi_S)^2\over d_A(\chi)^4}\int dl'l'^5  \nonumber \\
& &\times \int d\chi'{g(\chi',\chi)^2\over d_A(\chi')^8}P\left({l\over d_A(\chi)};\chi\right)P\left({l'\over d_A(\chi')};\chi'\right) \, ,
\eea
with
\be
\eta\equiv\left\{
\begin{array}{ll}
2 & \mbox{ for } \alpha\beta = \kappa\kappa		\\
3/2 & \mbox{ for } \alpha\beta = \kappa\epsilon	\\
1 & \mbox{ for } \alpha\beta = \epsilon\epsilon	\\
0 & \mbox{ otherwise }
\end{array}
\right.  .
\ee

Similarly, lens-lens coupling can be accounted for by iteratively expanding the recursive term in equation~\re{distort}.  
This yields the following second and third order source terms:
\bea
S^{(2L)}_{ab}&=&-2\Phi_{,ac}(\bmn d_A;\chi)\int d\chi'g(\chi',\chi)\Phi_{,cb}(\bmn d_A(\chi');\chi') \\
S^{(3L)}_{ab}&=&4\Phi_{,ac}(\bmn d_A;\chi)\int d\chi'g(\chi',\chi)\Phi_{,cd}(\bmn d_A(\chi');\chi')  \nonumber \\
& &\times \int d\chi''g(\chi'',\chi')\Phi_{,db}(\bmn d_A(\chi'');\chi'')
\, .
\eea
The first-second order correction to $C_{abcd}$ vanishes under the Limber approximation.  Cooray and Hu \cite{CH} 
calculated the second-second order correction, but neglected the correction involving the first and the third order  terms. This term does not
vanish; it can be written as
\bea
P^{(13)}_{abcd}&=&d_A(\chi)^{-6}\int d\chi'\int d^2n\,e^{i\bml\cdot\bmn}\VEV{S^{*(1)}_{ab}(\bmn d_A(\chi);\chi)S^{(3L)}_{cd}(\bmo;\chi')} \nonumber \\
& &+ (ab\leftrightarrow cd)\\
&=& 4d_A(\chi)^{-6}\int\!\!\int\!\!\int d\chi'd\chi''d\chi'''\int d^2n\,e^{i\bml\cdot\bmn} g(\chi'',\chi')g(\chi''',\chi'') \nonumber \\
& & \times \VEV{\Phi_{,ab}(\bmn d_A(\chi);\chi)\Phi_{,ce}(\bmo;\chi')\Phi_{,ef}(\bmo;\chi'')\Phi_{,fd}(\bmo;\chi''')} \nonumber \\
& & + (ab\leftrightarrow cd) \, .
\label{eq:P13b} 
\eea
To complete the derivation, each of the four $\Phi$\,s above must be expanded into its Fourier modes, $\phi({\bmk})$.  Then to calculate the ensemble average, $\VEV{\phi({\bmk})\phi({\bmk'})\phi({\bmk''})\phi({\bmk'''})}$, one must sum over all possible pairs of Wick contractions of the $\phi$\,s.  Contracting two $\phi$\,s ultimately produces delta functions which match up the corresponding $\chi$\,s.  For example, these contractions
$$
\langle
\contraction{}{\phi({\bmk})}{\phi({\bmk'})}{\phi({\bmk''})}
\contraction[2ex]{\phi({\bmk})}{\phi({\bmk'})}{\phi({\bmk''})}{\phi({\bmk'''})}
\phi({\bmk})\phi({\bmk'})\phi({\bmk''})\phi({\bmk'''})
\rangle
$$
lead to a term containing $\delta(\chi-\chi'')\delta(\chi'-\chi''')$.  In fact, this is the only non-trivial term: because of the weighting functions in \re{P13b}, delta functions in other terms ($\delta(\chi''-\chi')$ and $\delta(\chi''-\chi''')$) will cause the integrals to vanish \cite{CH}.  
Nonetheless, the first-third correction does not vanish completely, and we find that $P^{(13)}_{abcd}$ is coincidentally equal to $P^X_{abcd}$ in equation~\re{PX}.

There also exist corrections due to couplings between  Born and lens-lens source terms.  
However, in this case, we do not find additional corrections apart from those already considered 
by Cooray and Hu \cite{CH}. To the highest order in importance, i.e., fourth order
in potential perturbations, this completes the calculation of corrections to weak lensing angular power spectra.

\section{Results \& Summary}
\label{Sec:Results}

We have calculated the $O(\Phi^4)$ corrections to weak lensing power spectra that arise from the Born approximation and lens-lens coupling; they are illustrated in Fig.~\rf{graph}.  The primary effect of the new term, $C^X_l$, is to make the total correction to the $\kappa\kappa$, $\kappa\epsilon$ and $\epsilon\epsilon$ spectra negative for $l\ll 200$.  The $\beta$ and $\omega$ spectra are unaffected by the new term when comparing to the calculation of Ref.~\cite{CH}.   As discussed in \cite{CH}, 
the $\omega$ term arises from lens-lens coupling only; it is not a test of the Born approximation as was suggested in Ref.~\cite{Jain99}. The larger 
difference in the amplitude of the rotational power spectrum
shown here and the one shown in Ref.~\cite{HirSel03} (their Figure~2) is due to the large difference in the source redshift
used between these two calculations:  here we use $z_s=1$ for galaxy lensing surveys while $z_s=1100$ is used in \cite{HirSel03} for 
lensing of cosmic microwave background anisotropies. The large difference suggests that non-linear couplings that lead
to these second order corrections grow significantly with increasing path length between source and observer,
though the correction still remains below the first order term.
The conclusions of Cooray and Hu are unchanged: corrections 
related to the Born approximation and lens-lens coupling can be safely neglected for current surveys.  
They do not limit the use of $\beta$-modes in monitoring systematic errors. 

Furthermore, since our calculation is now complete, we can make even a stronger
statement. As shown in Figure~\rf{graph}, the higher order corrections are below the cosmic variance level of an all-sky survey, $\sqrt{2/(2l+1)}C_l$, 
out to a multipole of $10^4$ when these corrections start to be higher than the cosmic variance level.
This suggests that for all experiments that use statistical information out to a multipole of $10^4$, the angular power spectra will not be affected by the corrections associated with
the Born approximation and the coupling of two lenses. This, however, does not imply that theoretical estimates of the
angular power spectra are accurate to such a high precision. 
Throughout this paper, we have relied on the Limber approximation, which ignores contributions to the power spectra from modes with wave-vectors parallel to the line-of-sight.  This approximation is justified by lensing simulations \cite{WhiHu00}, 
which agree with the canonical calculation of the power spectra and hence our conclusions at the few percent level.
We have also assumed that the weak lensing calculation is valid, despite the fact that the 
lensing observables, such as convergence,
 have non-gaussian distributions with tails that can affect the non-linear part of their spectra, depending on how one includes high lensing peaks that merge weak lensing scales to strong lensing
in the data analysis \cite{ValeWhi03}. Beyond issues related to the weak lensing approximation,
the non-linear matter power spectrum for a given cosmology is uncertain at the 10\% level, both due to
uncertainties in the mapping between linear to non-linear dark matter fluctuations and the effect of baryons in determining how dark matter clusters on  arcminute scales and below \cite{White04,Zhan04}.
Accounting for such uncertainties are beyond those possible with analytical calculations. In fact, all remaining uncertainties are those that must be addressed with numerical simulations. 
We can  now be confident that the basic analytical formulation of lensing power spectra is well explored and the first order result is accurate enough for all upcoming experiments.

To summarize, here we have revisited the estimation of higher order corrections to the angular power spectra of weak gravitational lensing 
and have extended the  calculation of Cooray and Hu \cite{CH} that first described the corrections related to the
Born approximation and the neglect of line-of-sight coupling of two foreground lenses in the standard first order result. 
We found two additional terms to the fourth order in potential perturbations of the
large-scale structure and these terms, in return, altered
the convergence ($\kappa\kappa$),  the lensing shear E-mode 
($\epsilon\epsilon$), and their cross-correlation ($\kappa\epsilon$) power spectra on 
large angular scales. There is no modification to the
lensing shear B-mode ($\beta\beta$) and rotational ($\omega\omega$) component power spectra  when compared to the previous estimate.
With these new terms,  the calculation of corrections to weak lensing power spectrum  associated with both
the Born approximation and the lens-lens coupling is complete. The overall numerical result is that these
 corrections are unimportant for any weak lensing survey including for a full sky survey limited by the cosmic variance.

\begin{figure*}[!t]
\centerline{\psfig{file=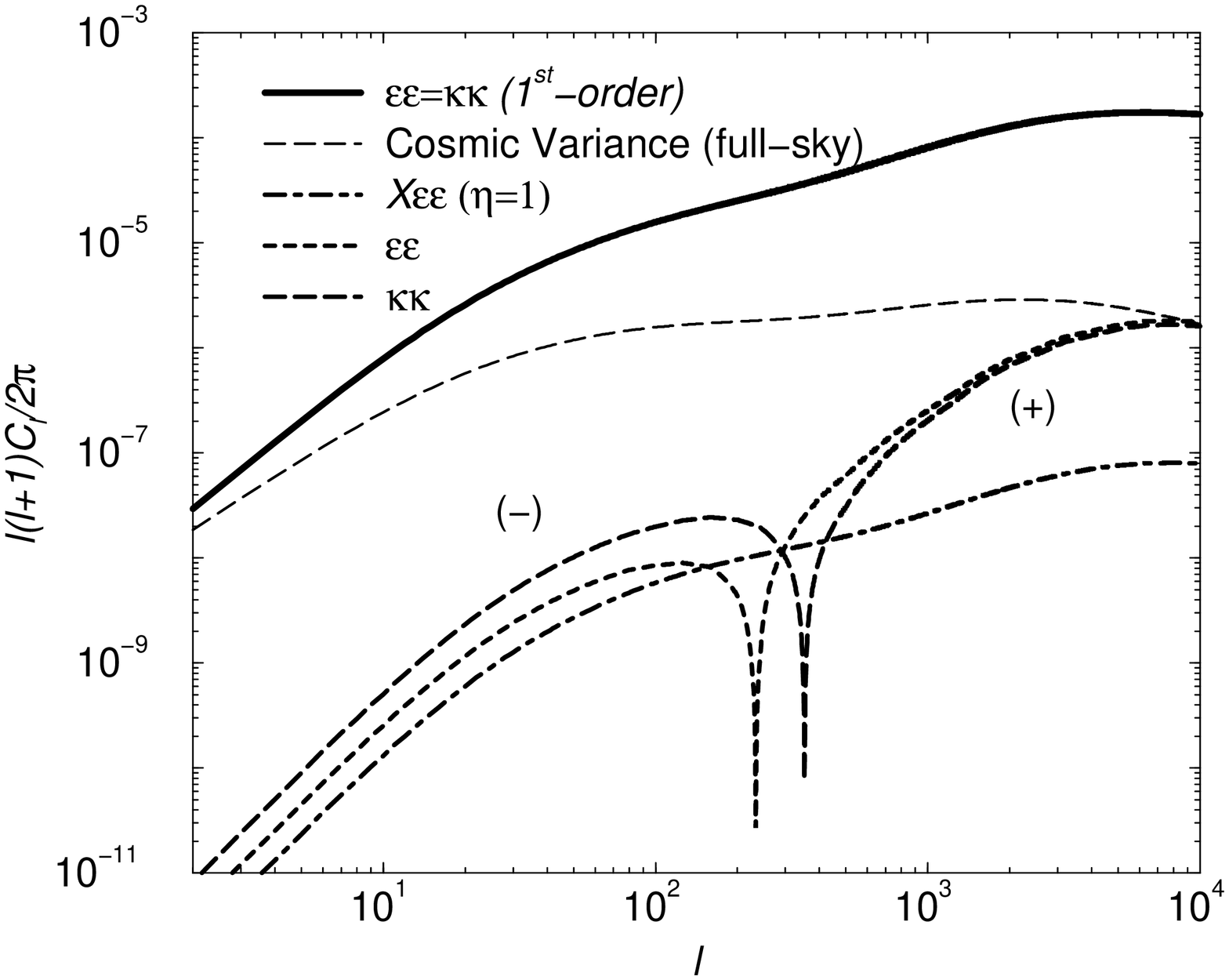,width=3.5in,angle=-90}
\psfig{file=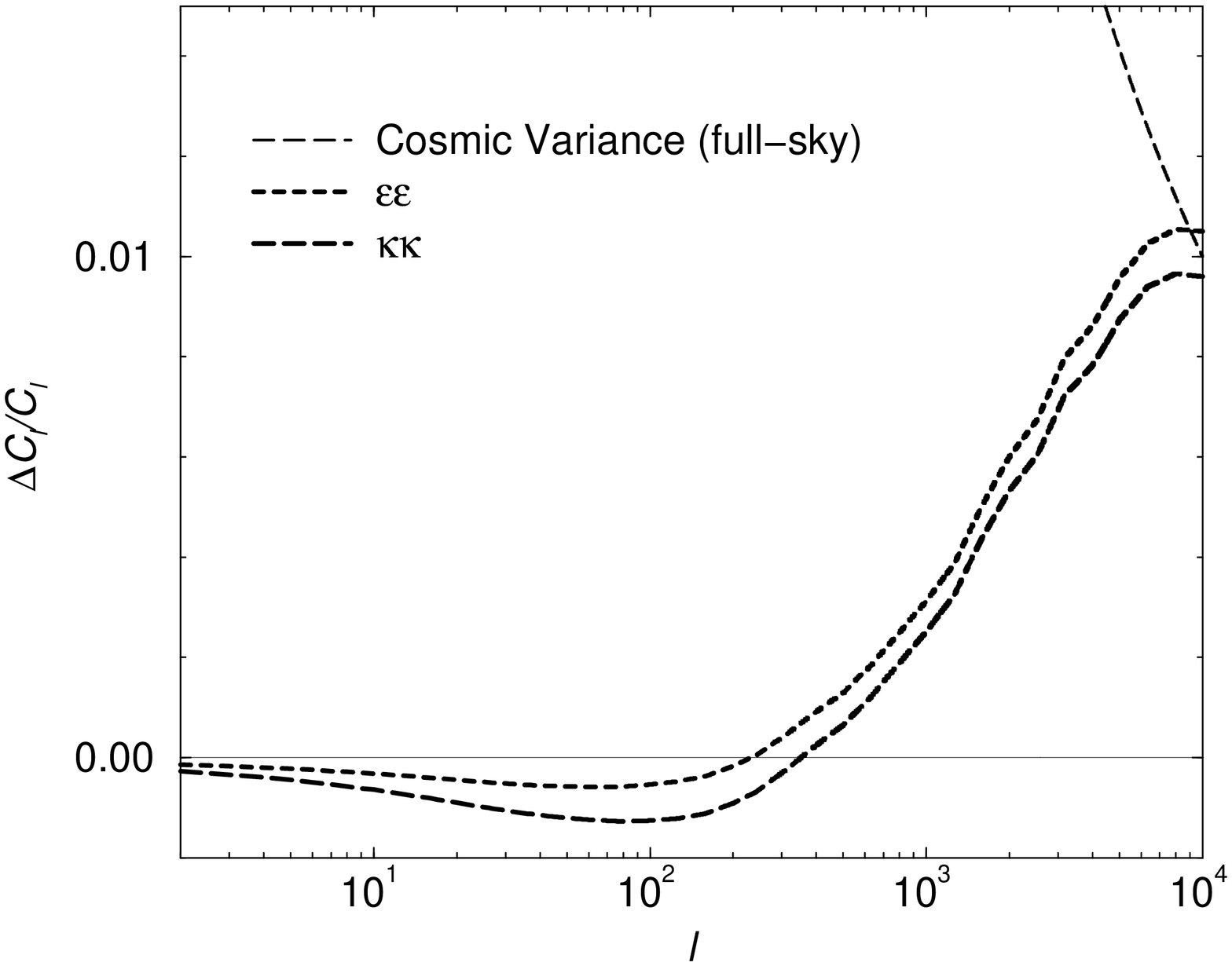,width=3.5in,angle=-90}}
  \caption{{\small Left: Weak lensing power spectrum and corrections.  The solid line shows the first order spectrum; the dot-dashed line shows the new correction, $C^{X\epsilon\epsilon}_l$ ($\eta=1$), which is negative; the dotted and dashed lines show the full corrections (Born and lens-lens) to the $\kappa\kappa$ and $\epsilon\epsilon$ spectra respectively -- they are negative at low $l$ and positive at high $l$ as indicated by (-)/(+) signs.  For reference, we also show the cosmic variance level of
an all-sky experiment with a thin long-dashed line. The comparison shows that for $l < 10^4$, the full corrections from Born and lens-lens coupling are
below the cosmic variance level and is unlikely to be an error of statistical significance.
Right: Full corrections divided by the first order result.  The dotted and dashed 
lines show the relative corrections for $\kappa\kappa$ and $\epsilon\epsilon$ respectively, while the thin long-dashed line is the ratio of cosmic variance to
the first order result.}}
  \label{fig:graph}
\end{figure*}

\section{Acknowledgments}
Authors thank Wayne Hu for helpful discussions, Martin White and Chris Vale for comments, and Scott Dodelson for inspiring this paper.  CAS was supported in part by the Department of Energy, the Kavli Institute for Cosmological
Physics at the University of Chicago and by NSF grant PHY-011442.

\end{document}